\documentclass[twocolumn,prb,aps,showpacs,superscriptaddress]{revtex4-1}
\usepackage{graphicx}
\usepackage{color}
\usepackage{dcolumn}
\usepackage{amsmath}

\newlength{\figwidth} 
 \newlength{\figwidthb} %
\begin{document}

\title{Quartz-based flat-crystal resonant inelastic x-ray scattering spectrometer with sub-10~meV energy resolution}

\author{Jungho~Kim}\email{jhkim@aps.anl.gov}
\affiliation{Advanced Photon Source, Argonne National Laboratory, Argonne, Illinois 60439, USA}
\author{D.~Casa}
\affiliation{Advanced Photon Source, Argonne National Laboratory, Argonne, Illinois 60439, USA}
\author{Ayman~Said}
\affiliation{Advanced Photon Source, Argonne National Laboratory, Argonne, Illinois 60439, USA}
\author{Richard~Krakora}
\affiliation{Advanced Photon Source, Argonne National Laboratory, Argonne, Illinois 60439, USA}
\author{B.~J.~Kim}
\affiliation{Department of Physics, Pohang University of Science and Technology, Pohang 790-784, Republic of Korea}
\affiliation{Max Planck Institute for Solid State Research, Heisenbergstra$\beta$e 1, D-70569 Stuttgart, Germany}
\author{Elina~Kasman}
\affiliation{Advanced Photon Source, Argonne National Laboratory, Argonne, Illinois 60439, USA}
\author{Xianrong~Huang}
\affiliation{Advanced Photon Source, Argonne National Laboratory, Argonne, Illinois 60439, USA}
\author{T.~Gog}\email{gog@aps.anl.gov}
\affiliation{Advanced Photon Source, Argonne National Laboratory, Argonne, Illinois 60439, USA}

\date{\today}

\begin{abstract}
Continued improvement of the energy resolution of resonant inelastic x-ray scattering (RIXS) spectrometers is crucial for fulfilling the potential of this technique in the study of electron dynamics in materials of fundamental and technological importance. In particular, RIXS is the only alternative tool to inelastic neutron scattering capable of providing fully momentum resolved information on dynamic spin structures of magnetic materials, but is limited to systems whose magnetic excitation energy scales are comparable to the energy resolution. The state-of-the-art spherical diced crystal analyzer optics provides energy resolution as good as 25~meV but has already reached its theoretical limit. Here, we demonstrate a novel sub-10~meV RIXS spectrometer based on flat-crystal optics at the Ir-L$_3$ absorption edge (11.215~keV) that achieves an analyzer energy resolution of 3.9~meV, very close to the theoretical value of 3.7~meV. In addition, the new spectrometer allows efficient polarization analysis without loss of energy resolution. The performance of the instrument is demonstrated using longitudinal acoustical and optical phonons in diamond, and magnon in Sr$_3$Ir$_2$O$_7$. The novel sub-10~meV RIXS spectrometer thus provides a window into magnetic materials with small energy scales.
\end{abstract}


\maketitle

\section*{Introduction}
The description of strongly correlated electron systems by means of elementary excitations has been a useful concept in modern condensed-matter physics~\cite{pines}. Magnetic excitations or magnons and their associated quasi-particles are traditionally studied with inelastic neutron scattering. It is a recent finding that resonant inelastic x-ray scattering (RIXS) can actually probe magnon as a result of the core-hole spin-orbit coupling of transition metal L edges~\cite{ament2011}. Nowadays, RIXS is routinely used to measure magnetic excitations in various different types of materials including cuprates~\cite{amentrmp}, iridates~\cite{jkim12,jkim13,jkim14} and osmates~\cite{calder2016}, with the highest energy resolution being $\sim$25~meV at the Ir L$_3$ absorption edge~\cite{jkim14,esrf13}. 

There are great interests in improving the energy resolution of the RIXS spectrometer to sub-10 meV level to fully extract information necessary to understand the low-lying magnetic dynamics of iridates which have been irresolvable within the current best energy resolution. Of particular recent interests are such elusive systems as quantum spin liquids, topological insulator, Axion insulator and Weyl semimetal, which are emergent quantum phases in correlated spin-orbit coupled materials~\cite{krempa14,pesin10,chaloupka10}.

The x-ray optics of state-of-the-art, high-resolution RIXS spectrometers are based on diced, spherical crystal analyzers in a near back-scattering geometry~\cite{gog13,yuri13}. Due to the resonant character of the technique, the incident energy is predetermined by the absorption edge of a given magnetic ion. However, the highly symmetric crystal structure of Si provides very few choices for near-backscattering reflections, thereby limiting the highest resolution achievable. Other crystals of lower symmetry such as quartz and sapphire in principle offer many more reflection choices, some of which have much smaller intrinsic energy width than Si with very good throughput~\cite{sutter05}. However, the technology to process these materials is still in its infancy~\cite{hasan17}.

In this article, a novel RIXS spectrometer with sub-10~meV energy resolution at the Ir L$_3$ edge is introduced, based on flat-quartz crystal optics. The unique advantage of the flat-crystal analyzer is that its resolution no longer depends on any geometrical factors, such as the focusing of the mirrors, the divergences of the X-ray beams, or the exact shape of the spherical analyzer. As a new important feature, efficient polarization analysis without loss of energy resolution is also demonstrated. The principle of the instrument design is described, the performance of its implementation is characterized and representative measurements of phonon dispersion in diamond and magnetic excitations in Sr$_3$Ir$_2$O$_7$ are provided. 

\begin{figure*}[ht]
\centering
\includegraphics[width=0.85\linewidth]{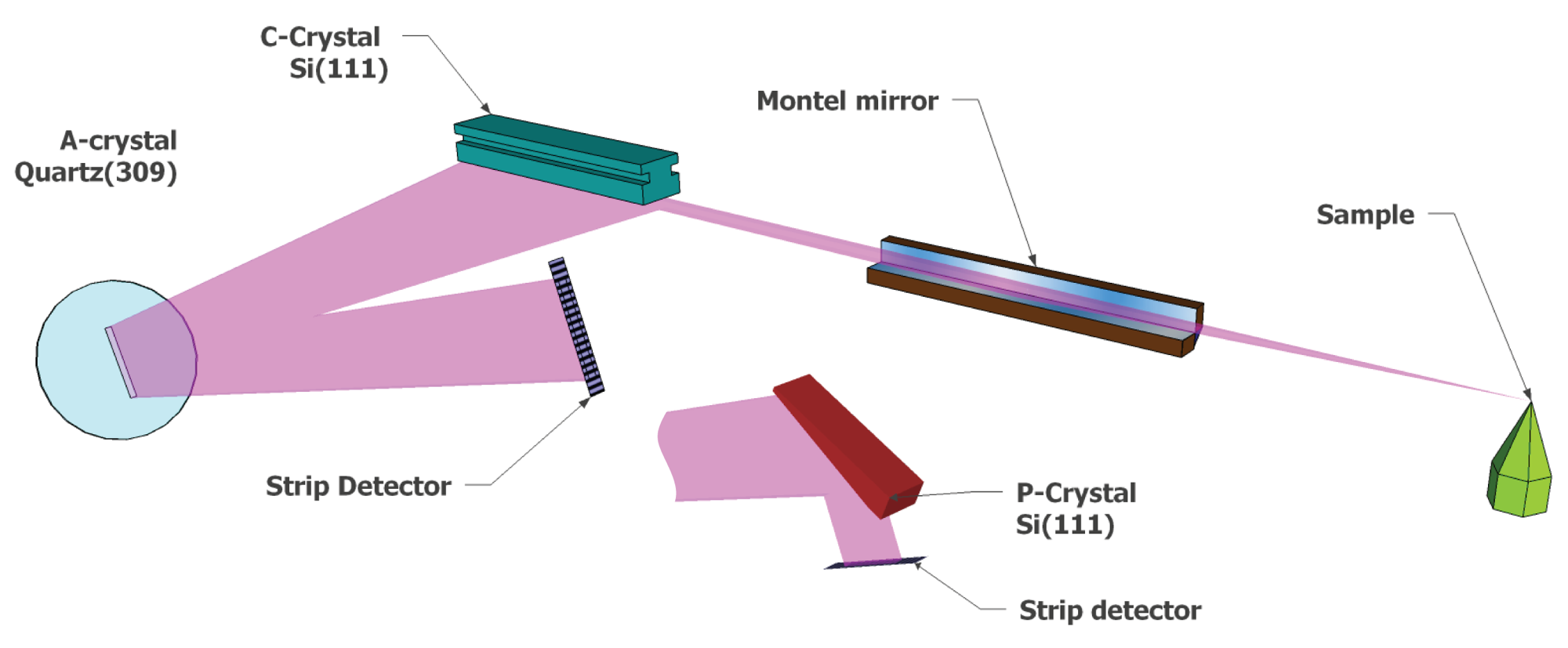}
\caption{Schematic of the flat-crystal RIXS spectrometer. This spectrometer, designed for the Ir L$_3$ edge at 11.215~keV, consists of a nested set (Montel) of two parabolic, laterally-graded Ru/C multilayer mirrors, followed by an asymmetrically cut Si(111) collimator C-crystal and a symmetric Quartz(309) analyzer A-crystal. The beam from the A-crystal is collected by a position-sensitive (strip-) detector. For polarization analysis, a Si(444) P-crystal is placed in the beam from the A-crystal, with the detector relocated to receive the diffracted beam from the P-crystal. }
\label{fig:schematic}
\end{figure*}

\section*{Experiment and Simulations}
The diamond(111) high-heat-load monochromator reflects x-rays from two in-line undulators into a four-bounce high-resolution monochromator, which consists of two monolithic Si(844) channel-cut crystals, resulting in an energy bandpass of 8.9~meV. The beam is then focused by a set of Kirkpatrick-Baez mirrors, yielding a typical spot size of 10$\times$40~$\mu$m$^2$ FWHM (v$\times$h) at the sample. The sample scattering plane is horizontal, while the scattering plane of the CA crystals is vertical. 

A laterally-graded parabolic collimating nested (Montel) mirror, where two one-dimensionally figured multilayer mirrors are attached orthogonally, was chosen for its superior surface figuring, compact design, and easy instrumentation and stability control, compared with traditional designs such as a Kirkpatrick-Baez (KB) system. It is designed to collimate scattered x-ray beams around the Ir-L$_3$ absorption photon energy (11.215~keV) in both vertical and horizontal directions. Mirror parameters (parabolic shape, overall dimensions, choices of multilayer reflector and spacer elements) were determined by the input requirements of the CA RIXS spectrometer. The distance between the scattering source and the mirror center is 200~mm and the mirror length is 150~mm. The Bragg angle is 1.02$^{\circ}$ and the angular acceptance is as large as 14.5~mrad. The multilayer reflector and spacer are Ru and C, respectively. Detailed design parameters and its performance can be found in Ref.~25. A small vertical spot size is important to obtain the required x-ray collimation ($<$100~$\mu$rad) after the Montel mirror. Previous study showed that the x-ray collimation in the horizontal direction is rather poor ($\sim$200~$\mu$rad) due to the large horizontal spot size~\cite{jkim16}. However, this does not significantly affect the efficiency or energy resolution of the CA analyzer because the horizontal angular acceptances of both C and A crystals are larger than 200~$\mu$rad. 

In designing the CA-RIXS spectrometer, care was taken to maximize the incident solid-angle acceptance, optimize the throughput, maintain the best energy resolution and provide efficient polarization analysis without sacrificing resolution. This was accomplished by carefully selecting suitable crystal reflection and asymmetry angles. For guidance in this selection process, diffraction conditions and crystal scans were simulated in extended 3$-$dimensional DuMond diagrams,~\cite{dumond} based on two-beam dynamical diffraction theory~\cite{authier} and representing intensities after successive reflections and asymmetry transformation as a function of angle and energy.

\section*{Principle of the flat crystal RIXS analyzer}
Flat-crystal x-ray optics are superior in performance to curved crystal systems, because they are free from figure errors and strains that degrade the performance~\cite{dickinson08,bortel00,yuri14,huang11}. However, the angular acceptance of flat crystal systems is so small that they are not practical in situations where weak, radially scattered radiations from a sample needs to be collected, as is the case of a RIXS spectrometer. Recent advances in multilayer optics provide a way to overcome this impediment~\cite{cai13,honniche11,mundboth14,suvorov14} through the use of curved, laterally-graded mirrors, that can easily accept radiations in a large solid angle.

\begin{figure}[b]
\centering
\includegraphics[width=\linewidth]{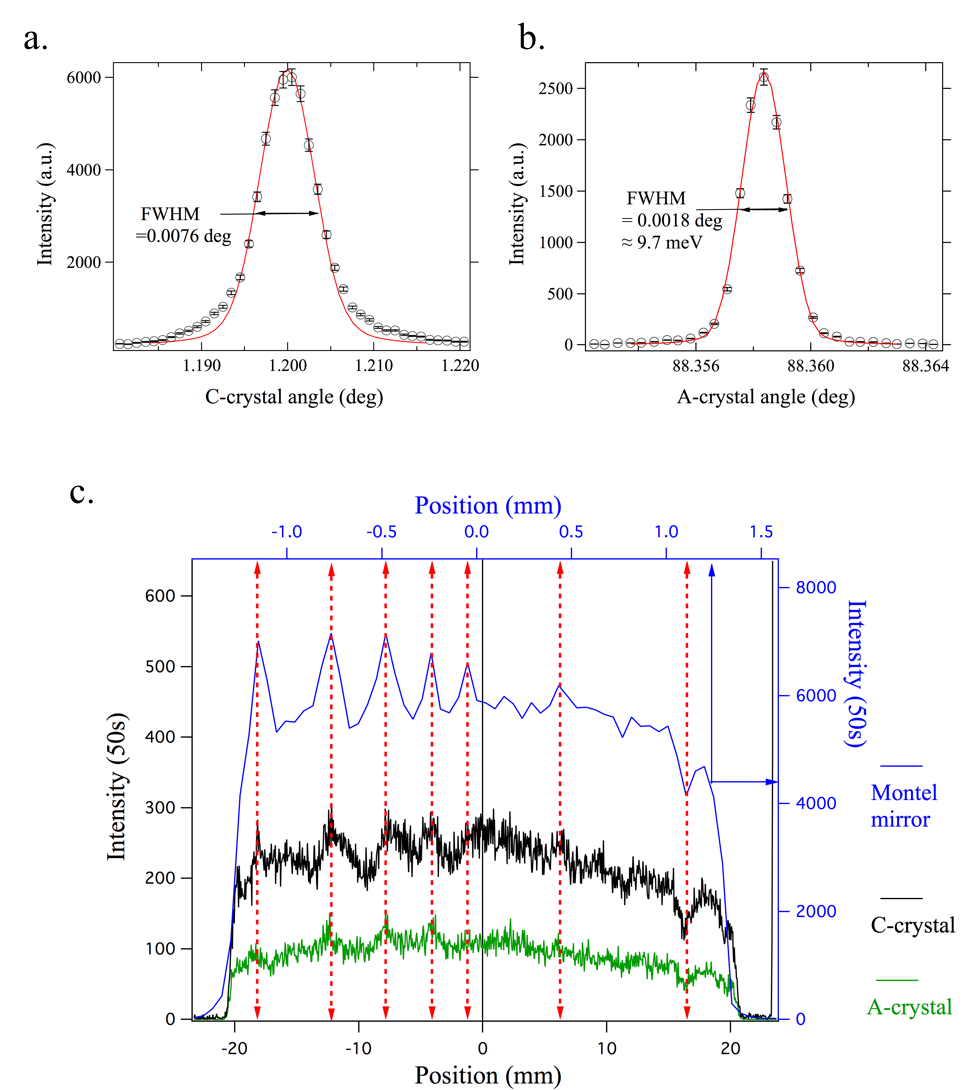}
\caption{Measured angle scans and x-ray intensity profiles. (a) The FWHM of the C-crystal angle scan is ~0.0076$^{\circ}$, which agrees with simulations based on its asymmetry angle and assuming 92~$\mu$rad incident divergence from the Montel mirror. (b) The FWHM of the A-crystal scan is ~0.0018$^{\circ}$, in agreement with simulations based on a divergence of 6.1~$\mu$rad from the C-crystal and a 8.9~meV bandpass. This angular width corresponds to an energy spread of 9.7~meV. (c) X-ray intensity profiles as measured by the positon-sensitive detector. Blue trace, top-axis, corresponds to downstream of  Montel mirror, while black and green traces, bottom axis, correspond to downstream of the C- and A-crystal respectively. The width of the blue profile indicates that the degree of collimation by the Montel mirror is as expected. The preservation of the same distinct pattern before and after the C-crystal indicates that collimation by the C crystal performs as designed.}
\label{fig:fig2}
\end{figure}

Figure~\ref{fig:schematic} shows a prototype of the flat crystal RIXS analyzer implemented for the operation at the Ir L$_3$ absorption edge (11.215~keV) at the RIXS beamline 27-ID of the Advanced Photon Source. A nested set of two parabolic, laterally-graded Ru/C multilayer mirrors, called the Montel mirror, is located at a distance of 200~mm from the sample to serve as initial collimator~\cite{jkim16}. It accepts more than 10$\times$10~mrad$^2$ of scattered radiation from the sample in the vertical and horizontal directions, and collimates it to less than 100$\times$100~$\mu$rad$^2$. A Si(111) asymmetrically cut C-crystal whose angular acceptance matches the emittance of the Montel mirror is positioned at $\sim$500~mm downstream, and it further reduces the beam divergence to $\sim$6~$\mu$rad in the vertical direction. This is smaller than the acceptance of the A-crystal that is placed at a distance of $\sim$1000~mm from the C-crystal. For the A-crystal, the availability of nearly perfect $\alpha$-Quartz provides a near-backscattering reflection (309) with an intrinsic energy width of 3.7~meV, $\sim$4 times smaller than the highest resolving Si(844) reflection (14.6~meV) at this incident energy. It should be emphasized that the quality of collimation only affects the efficiency of the CA-analyzer, while its energy resolution remains solely determined by the intrinsic width of the A-crystal. The X-ray beam from the A-crystal is collected by a position-sensitive detector (Mythen by Dectris), with an effective length of 64~mm, divided into 1280 pixels of 50~$\mu$m width each. In the following, this RIXS analyzer system will be referred to as a CA-analyzer.

In an ideal RIXS measurement, the energy, momentum and polarization of scattered radiation are measured. To date, polarization measurements using curved crystal optics have been inefficient and tend to compromise the energy resolution~\cite{ishii11,gao16}. A remarkable, additional advantage of the CA analyzer is to analyze polarization with high efficiency and without loss of energy resolution. For this purpose, the beam from the A crystal is further diffracted by a Si(444) polarizer P-crystal at a Bragg angle near 45$^{\circ}$, with the detector relocated to receive the beam from the P-crystal.

\begin{figure}[b]
\centering
\includegraphics[width=\linewidth]{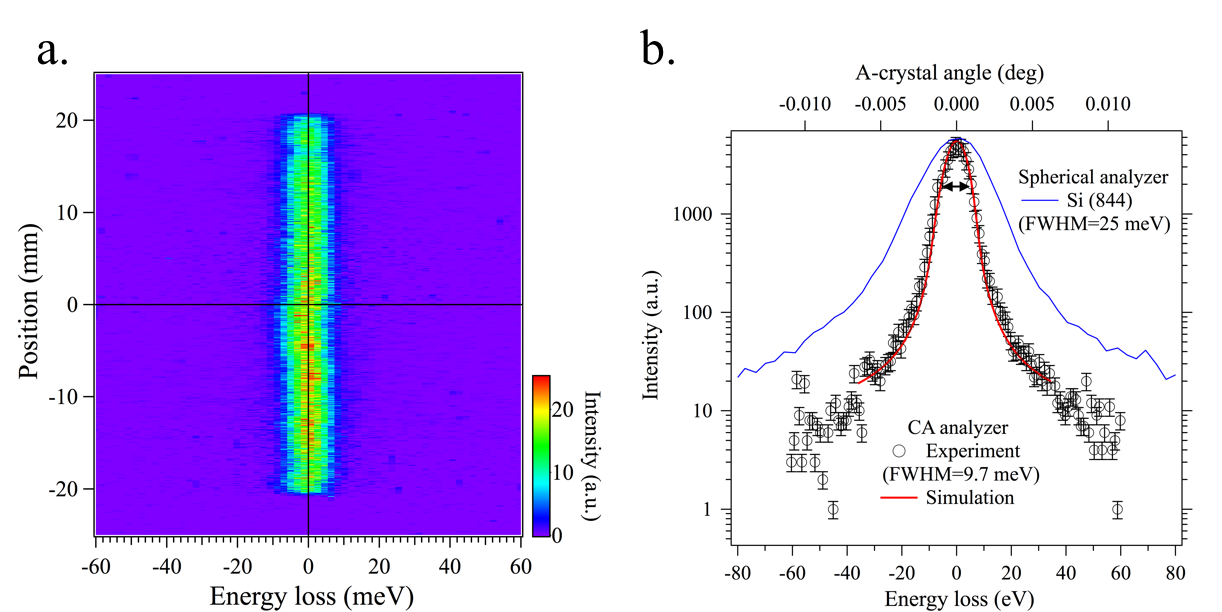}
\caption{Spectral resolution functions. (a) Image plot of elastic scattering as a function of incident energy and position along the detector. (b) Spectral resolution function of the flat-crystal spectrometer (open circles), as obtained by integrating intensity along position. FWHM is ~9.7~meV. Its profile agrees well with simulation (red line). For comparison, the resolution function of a Si(844) spherical analyzer (25~meV) (blue line) is shown.}
\label{fig:fig3}
\end{figure}

\begin{figure*}[ht]
\centering
\includegraphics[width=0.85\linewidth]{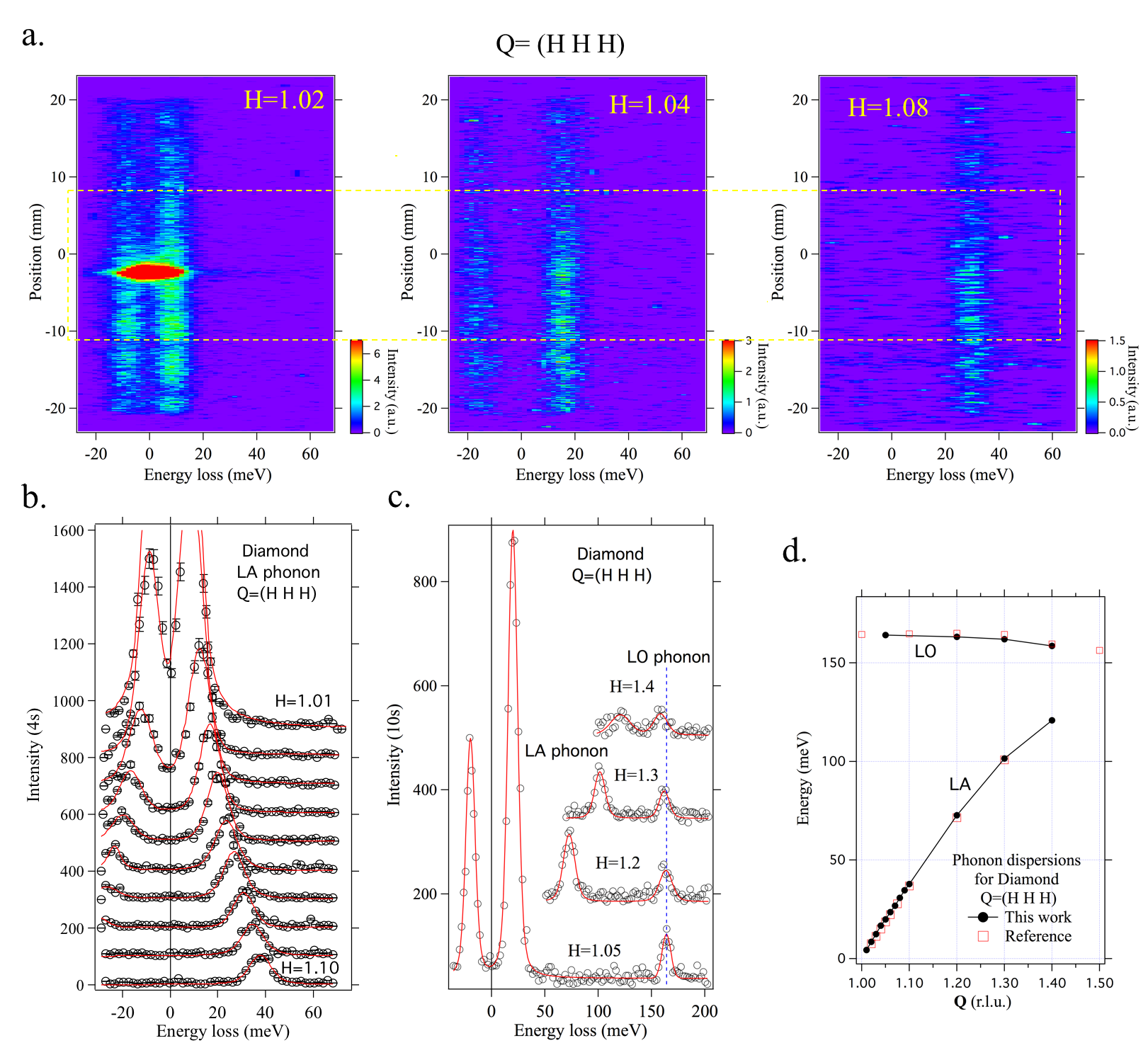}
\caption{Longitudinal acoustic (LA) and optical (LO) phonons in diamond. (a) Image plots of the LA phonon spectra for H=1.02 (left), 1.04 (center), and 1.08 (right). The two streaks at positive and negative energy loss correspond to Stokes and anti-Stokes scattering respectively. The strong intensity at zero energy loss for H=1.02 comes from the (1 1 1) Bragg peak. The yellow dashed lines indicate the integration range for extracting phonon spectra, restricted for better momentum resolution. (b) Phonon spectra for H=1.01-1.10. (c) Phonon spectra for H=1.05, 1.2, 1.3, and 1.4 with extended energy range. The LO phonons are now visible around 160~meV. d. Dispersion curves for both longitudinal modes. Measured data agree well with IXS results.}
\label{fig:fig4}
\end{figure*}

\section*{Results}
\subsection*{Performance of the flat-crystal RIXS spectrometer}
For alignment and characterization, elastically scattered radiation from a tape sample (3M Scotch Magic tape) is used. The incident beam is monochromatized to a band pass of 8.9~meV. The CA analyzer is positioned at a scattering angle 2$\theta$ = 15$^{\circ}$. In order to assess the performance of the CA-analyzer, static diffraction characteristics as well as dynamic rocking curves and energy scans were extensively simulated. Extended DuMond diagrams~\cite{dumond} based on two-beam dynamical diffraction theory~\cite{authier} were used to compute 3-dimensional reflectivity profiles as function of both the angle and the energy at various points along the chain of the crystals. In order to calculate the energy- or angle-scan profiles, the corresponding profiles are convoluted along the angle- or energy-axis. 

Figure~\ref{fig:fig2}a shows the measured rocking curve of the asymmetric Si (111) C-crystal. The measured full-width-half-maximum (FWHM) value is 0.0076$^{\circ}$, consistent with the simulated results (solid curve) that are based on a model with an incidence divergence of 92~$\mu$rad from the mirror and an incident energy bandwidth of 8.9~meV. Figure~\ref{fig:fig2}b shows the rocking curve for the symmetric quartz (309) A-crystal. The FWHM is about 0.0018$^{\circ}$, again consistent with the simulations. These results indicate the near-ideal quality of the C- and A-crystals, as well as their strain-free mountings and an effective degree of collimation of the Montel mirror.   

Regarding the measured efficiency, the C-crystal reflects ~70~$\%$ of x-rays coming from the Montel mirror, while the A-crystal passes ~30~$\%$  of x-rays from the C-crystal to the detector, amounting to an overall efficiency of about 21~$\%$ . This is in very good agreement with the simulations that give an overall efficiency of 22~$\%$, which indicates that the successive collimation by the Montel mirror and the C-crystal works exactly as designed. This is also borne out by more detailed observations of the vertical beam profiles measured at various points along the beam path using the position-sensitive detector. Figure~\ref{fig:fig2}c shows the profiles downstream of the Montel mirror, the C-crystal and the A-crystal. Firstly, the full size of the beam emerging from the Montel mirror was measured to be about 2.6~mm as shown in the blue line (top and right axes). The angular acceptance corresponding to this measured size is about 14.5~mrad. The intensity modulations are attributed to the figure errors of the Montel mirror. The black trace in Fig.~\ref{fig:fig2}c shows the beam profile downstream of the C-crystal. The measured beam size is now 40.8~mm, consistent with the designed asymmetry factor of b = 15.6, indicating that the full beam from the Montel mirror is accepted and passed along. The incident intensity fluctuations are preserved. Finally, the green trace shows the beam profile downstream of the symmetric A-crystal. Again, the intensity fluctuations are preserved, and the reduction in intensity is consistent with a reduced reflectivity for quartz and its narrow energy acceptance.

The spectral resolution function of the CA-analyzer can be obtained in two ways: by scanning the analyzer itself or by scanning the incident energy. For the first scan mode, a single angle scan of the A-crystal is sufficient, because the energy bandpass of the C crystal is very large (~6.3 eV) compared to that of the A-crystal. Thus, the A-crystal rocking curve shown in Fig.~\ref{fig:fig2}b, when the angle is converted to the energy, gives the spectral resolution function with a width of just under 10~meV. For the second mode, the spectral resolution function is obtained by scanning the high-resolution monochromator. Figure ~\ref{fig:fig3}a shows an image plot of such an incident energy scan. The x-axis represents the energy loss, defined as the difference between the incident and scattered energy. The y-axis is the position on the position-sensitive detector. It can be seen that all elastic intensities are well aligned along the zero-energy loss. The spectral resolution function is obtained by integrating spectral intensities at all the positions as a function of energy loss. Figure~\ref{fig:fig3}b shows the spectral resolution function obtained in this way. It has a FWHM of 9.7~meV, consistent with the corresponding simulation (red solid curve) and the result from Fig.~\ref{fig:fig2}b. Considering the energy bandpass of the incident beam of 8.9~meV, the intrinsic energy resolution of the CA-analyzer can be approximated by $\sqrt{9.72^2-8.92^2}$ = 3.9~meV, very close to the theoretical value of 3.7~meV. For comparison, the spectral resolution function of a spherical diced Si(844) analyzer is also plotted.

\subsection*{Longitudinal acoustic and optical phonons in Diamond crystal}
For the purpose of characterizing the performance of the flat-crystal RIXS spectrometer and calibrating its energy scale, we have reproduced the well-known phonon spectra from diamond, which were investigated previously by various high resolution spectroscopies such as inelastic neutron scattering (INS), Raman scattering, and inelastic x-ray scattering (IXS)~\cite{verbeni08}. Diamond phonons are well suited for the purpose because their energies span a sufficiently large energy window, and the specific balance factor precisely defines the intensity ratio between Stokes and anti-Stokes scattering, allowing judgment of the uniform response of the instrument.

For these measurements, a flat Diamond(111) crystal was used with the azimuthal angle chosen such that the scattering (1 0 -1) direction was contained in the scattering plane. Longitudinal acoustic (LA) and optical (LO) phonons were measured along $\bf{Q}$=(H H H), with 1 $<$H$<$1.5. The momentum resolution corresponding to the angular acceptance of the mirror of 14.5 mrad is about 0.022 rlu, or 0.39 nm$^{-1}$.

\begin{figure}[t]
\centering
\includegraphics[width=\linewidth]{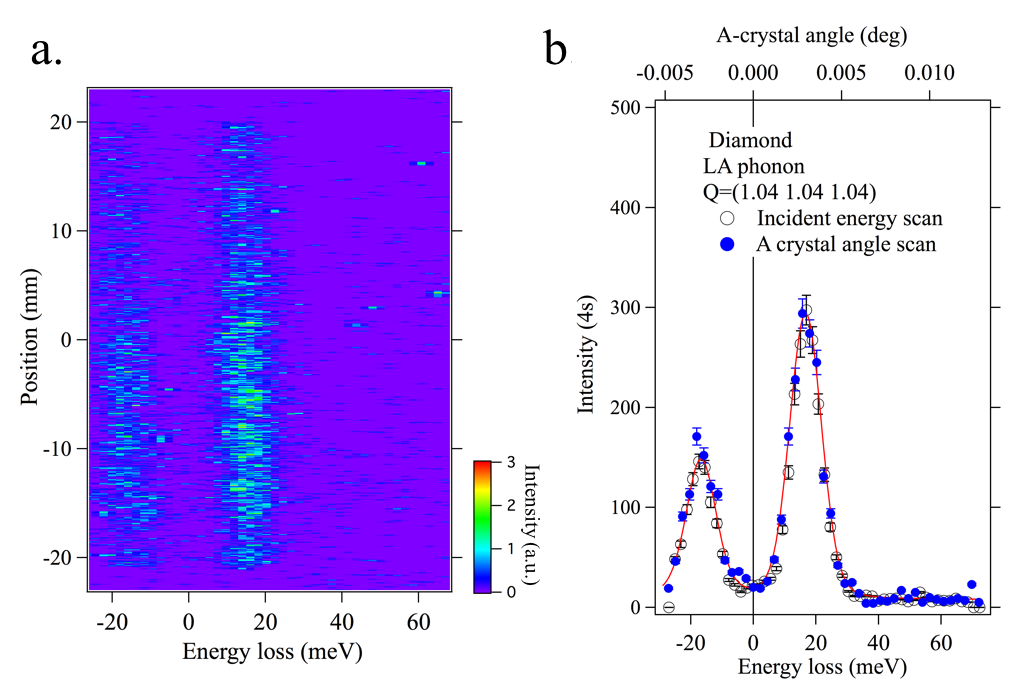}
\caption{Phonon spectrum for H=1.04 obtained by A-crystal angle scan. a. Image plot of the A-crystal angle scan, equivalent to the center panel of Fig.~\ref{fig:fig4}a and ~\ref{fig:fig4}b. The phonon spectrum obtained with both scan modes agree well with each other. The red line is a Voigt fit.}
\label{fig:fig5}
\end{figure}

The image plot on the left of Fig.~\ref{fig:fig4}a shows the LA phonon spectrum at H=1.02 recorded by the CA-analyzer. The strong intensity near zero energy loss is from the (1 1 1) Bragg peak. The two intensity streaks at positive and negative energy loss correspond to Stokes and anti-Stokes scattering, respectively. The image plot in the center shows the LA phonon spectrum at H=1.04. Here, elastic scatterings are suppressed and the separation between the Stokes and anti-Stokes scattering becomes larger. For H=1.08, as shown in the image plot on the right, anti-Stokes scattering is out of the measured range and the Stokes scattering appears at a higher energy loss. The phonon spectra in Fig.~\ref{fig:fig4}b and ~\ref{fig:fig4}c were obtained by integrating the intensities as a function of the energy loss. Taking advantage of the position-sensitive detector, a smaller momentum resolution (0.011 rlu) was selected by restricting the integration position range to half the length of the detector (20 mm). This is indicated by the yellow dashed lines in Fig.~\ref{fig:fig4}a. For the case of H=1.01, 1.02, and 1.03, the areas with strong elastic intensity around the $-$2~mm position are excluded. This is one of the powerful features of the flat-crystal RIXS spectrometer in combination with a position-sensitive detector; the integration ranges can be chosen a posteriori to set the momentum resolution or mask off unwanted areas. 

Figure~\ref{fig:fig4}b shows the LA phonon spectra. The measured spectra are fitted using the Voigt functions that satisfy the detailed balance factor. Figure~\ref{fig:fig4}c shows the spectra for H=1.05, 1.2, 1.3, and 1.4 over a wider energy range. The LO phonon is seen at around 160~meV, along with the highly dispersing LA phonon. Fig.~\ref{fig:fig4}d shows the resulting dispersion of the two longitudinal phonon modes. The results from the CA-analyzer agree very well with the IXS results, indicating that the high-resolution monochromator is properly calibrated. Alternatively, the same spectra can be obtained by scanning the angle of the A-crystal. Figure~\ref{fig:fig5}a shows an image plot for H=1.04 taken by such an angle scan, which is directly comparable to the center plot in Fig.~\ref{fig:fig4}a. The phonon spectra obtained in both scan modes shown in Fig.~\ref{fig:fig5}b agree well with each other. 
\begin{figure}[t]
\centering
\includegraphics[width=0.6\linewidth]{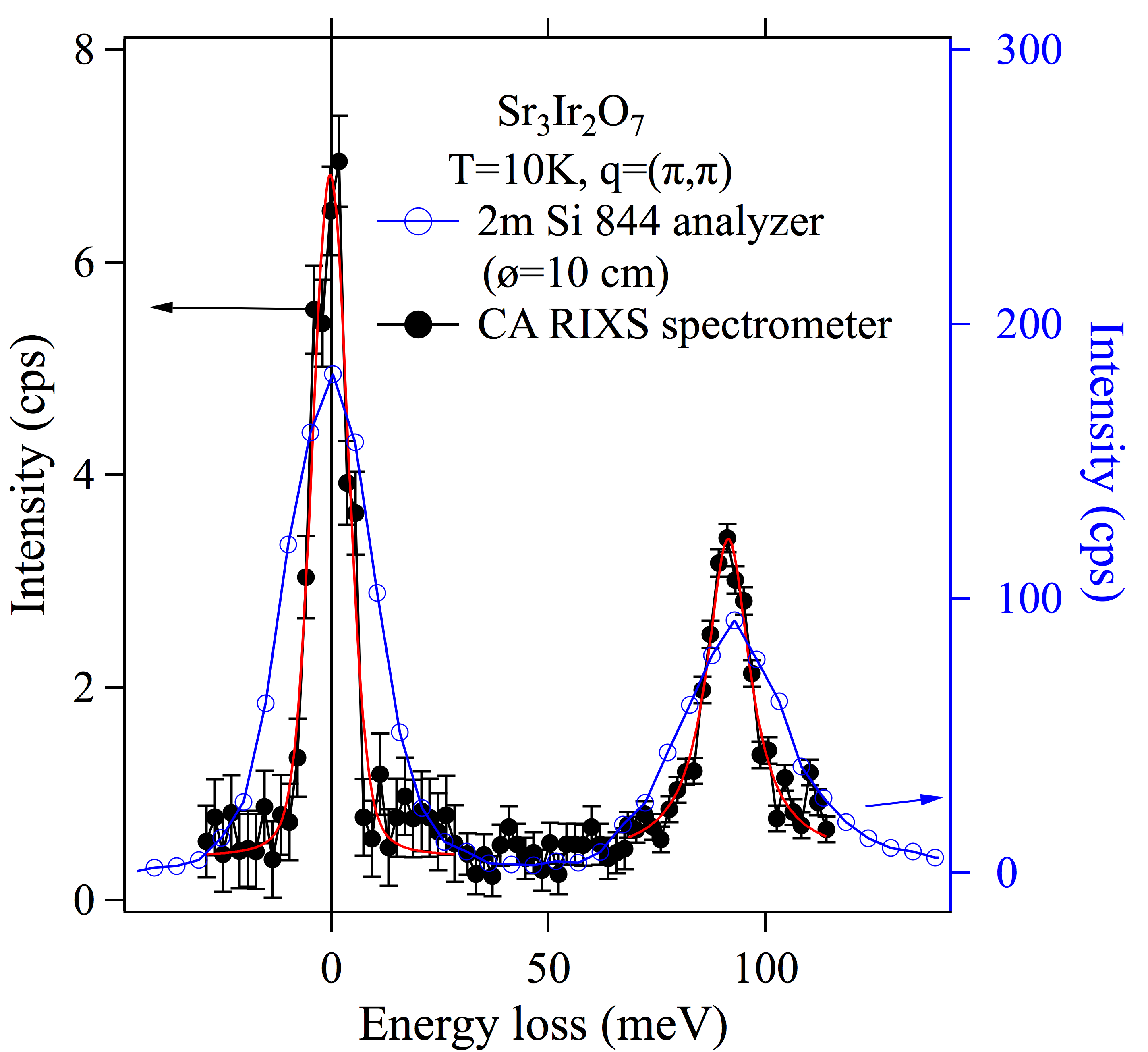}
\caption{The magnon spectrum of Sr$_3$Ir$_2$O$_7$ at the magnetic zone center. The spectrum measured by the flat-crystal RIXS spectrometer is shown as filled black circles (left axis). The elastic line indicates the overall energy resolution of 9.7~meV. A sharp magnon feature is seen at $\sim$90~meV. The red line is the Voigt fit. For comparison, a spectrum obtained with a standard RIXS spectrometer (25~meV resolution) is also shown (open blue circles, right axis).}
\label{fig:fig6}
\end{figure}
\subsection*{Magnon spectrum in Sr$_3$Ir$_2$O$_7$ at the magnetic zone center }
The bilayer iridate Sr$_3$Ir$_2$O$_7$ is a magnetic insulator driven by spin-orbit coupling with a small charge gap~\cite{moon08}. Previous RIXS studies have revealed an exceptionally large magnon gap of $\sim$90~meV resulting from bond-directional, pseudodipolar interactions that are strongly enhanced near the metal-insulator transition boundary~\cite{jkim13}. Figure~\ref{fig:fig6} shows the magnon spectrum in Sr$_3$Ir$_2$O$_7$ at the magnetic zone center as measured by the CA-RIXS spectrometer. The elastic scattering at zero energy loss has a FWHM of 9.7~meV, confirming the overall energy resolution determined earlier. A sharp magnon feature is seen at $\sim$90~meV, with a FWHM of 12~meV as determined from a Voigt profile fit. Magnon peak measured by a standard RIXS spectrometer with an overall energy resolution of 25~meV~\cite{jkim13} is also plotted in Fig.~\ref{fig:fig6} for comparison.

\subsection*{Polarization analysis with high efficiency and high energy resolution}
Scattered radiation typically contains both $\sigma-$ and $\pi-$polarizations~\cite{amentrmp}. Equally efficient detection of both polarizations is essential. The CA-analyzer presented in this work has a calculated $\pi/\sigma$ detection ratio of 0.88, suitable for a practical instrument.

\begin{figure}[t]
\centering
\includegraphics[width=\linewidth]{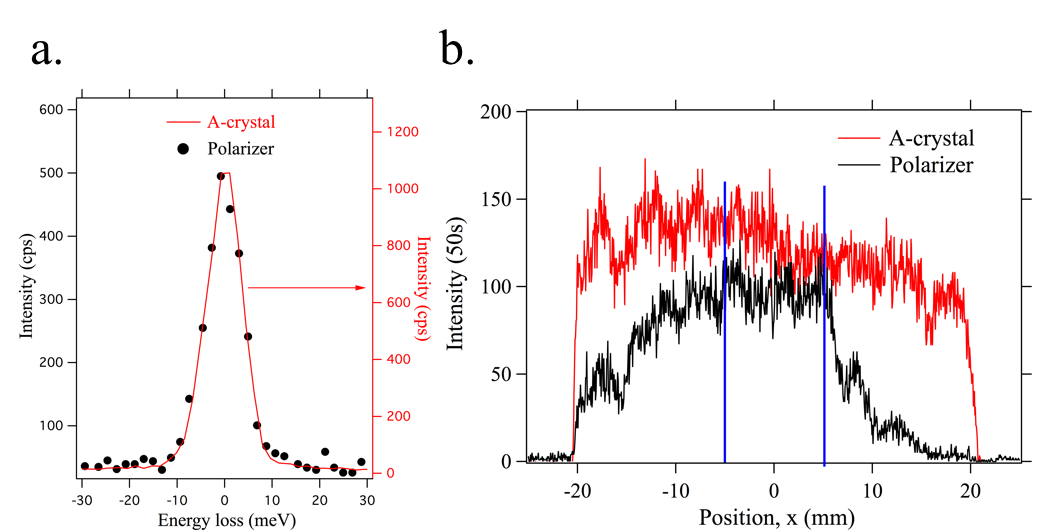}
\caption{Polarization analysis performance (polarizer setup as shown in Fig.~\ref{fig:schematic}) (a) Elastic spectra at the scattering angle of 2$\theta$= 15$^{\circ}$ with the polarizer and without the polarizer, obtained by scanning the incident energy. Two spectra are virtually indistinguishable from each other, with a FWHM of 9.7~meV. (b) Intensity profiles of the A- and P-crystal as a function of position along the detector. The low overall efficiency is due to poor reflectivity of the P-crystal along its length, except for the region around the center ($-$5~mm$<$x$<$5~mm) indicated by the blue lines.}
\label{fig:fig7}
\end{figure}

Polarization analysis can then be accomplished by including a polarizer between the A-crystal and the detector. A polarizer (P) crystal should have a scattering angle close to 90$^{\circ}$. Si(444) provides 89.68$^{\circ}$ at 11.215keV, and is therefore suitable. A symmetric Si(844) crystal was used for this proof-of principle study. As shown in Fig.~\ref{fig:schematic}, the P-crystal is placed in the beam emerging from the A-crystal and the detector is relocated to intercept the diffracted beam from the P-crystal.

In order to determine the energy resolution in the polarization analyzing mode, elastic scatterings from a tape sample were collected at scattering angles of 2$\theta$=0$^{\circ}$ and 13$^{\circ}$. Figure~\ref{fig:fig7}a shows the spectra for both scattering angles, obtained by scanning the incident energy. Two spectra are virtually indistinguishable from each other, with a FWHM of 9.7~meV. These measurements clearly show that the overall energy resolution is unaffected by adding a polarization analyzer.  

Regarding the efficiency, however, the P-crystal used for this demonstration was found to be somewhat strained, not diffracting uniformly over its length and rendering the measured overall efficiency (35$\%$) lower than expected. Figure~\ref{fig:fig7}b shows the intensity profiles of the A- and P-crystal as a function of position along the detector, recorded at peak intensity. It can be seen that the low overall efficiency is due to poor reflectivity of the P-crystal along its length, except for a small region around the center ($-$5~mm$<$x$<$5~mm). Considering only this region, the reflection efficiency is as good as 76~$\%$, comparable to what was expected from simulations.

\section*{Discussion and Outlook}
In this work, a novel, flat-crystal RIXS spectrometer has been designed and implemented, achieving an unprecedented energy resolution better than 10~meV at the Ir-L3 edge (with the resolution of the CA-analyzer being only 3.9~meV), unprecedented for any RIXS measurement. Besides its superior energy resolution, the flat-crystal system has two additional advantages: the capability to analyze polarization with no loss of resolution and good efficiency, and convenient selection of the momentum resolution.

Improvements in performance and extensions to other absorption edge energies can be realized in future implementation of the instrument. The best energy resolution at optimal throughput is achieved when the incident band pass is matched to the intrinsic energy width of the A-crystal. For the present prototype, according to simulations, reducing the incident energy bandpass from the current 8.9~meV to 3.7~meV would improve the overall energy resolution from 9.7 to 5.2~meV (with the same throughput) and raise the spectral efficiency from 21$\%$ to 49$\%$. 

In order to extend the concept of the flat-crystal spectrometer to other incident energies, its components should be selected following these guidelines: (1) The choice of A-crystal determines the ultimate energy resolution and therefore a reflection with an appropriately small intrinsic width and high reflectivity should be selected. Compared with silicon that has only a few tens of Bragg reflections in the medium energy range ($<$16~keV), quartz has hundreds to thousands of reflections. This enables us to implement flat-quartz analyzers at virtually any photon energies with resolution up to 1~meV. (2) The C-crystal should be of low reflection order, to attain a large angular acceptance at moderate asymmetry. Si(111) is a good choice for most situations. The Montel mirror needs to be designed to achieve appropriate collection, collimation and throughput at a chosen working distance. (3) Finally, the P-crystal needs to provide a near 90$^{\circ}$ scattering angle at the desired energy, have good reflectivity, an asymmetry chosen so that the full beam from the A-crystal is accepted, and the diffracted beam remains wholly within the detector length.

\acknowledgements{The use of the Advanced Photon Source at the Argonne National Laboratory was supported by the U.S. DOE under Contract No. DE-AC02-06CH11357.}

\bibliography{carixs}

\end{document}